\documentclass[10pt,twocolumn,twoside]{IEEEtran}

\IEEEoverridecommandlockouts                            
\usepackage{myStyleIEEE}
\usetikzlibrary{backgrounds}
\usepackage{nicefrac}
\usepackage{microtype}
\addshortterm{n}{GVL}{Generalized Vector Locus}{}
\addshortterm{n}{RRF}{Reduced Reference Frame}{}

\renewcommand\rev[1]{#1} 
\newcommand\update[1]{{#1}} 
\newcommand\secondupdate[1]{{#1}} 
\newcommand\camera[1]{\textcolor{blue}{#1}}
\renewcommand\camera[1]{#1}

\newif\ifshowauthors\showauthorstrue

\usepackage{graphicx}
\usepackage{textcomp}
\usepackage{balance}

\usepackage{lipsum}

\title{\LARGE\bf 
Generalized Vector Locus Transformation\\
for Unbalanced Three-Phase Systems}

\ifshowauthors
\author{Maitraya~Avadhut~Desai\,\orcidlink{0009-0000-1383-0773},~\IEEEmembership{Graduate~Student~Member,~IEEE,}
        Francisco~Escobar\,\orcidlink{0000-0002-6235-5689},~\IEEEmembership{Member,~IEEE,}
        Gabriela~Hug\,\orcidlink{0000-0002-4312-616X},~\IEEEmembership{Senior~Member,~IEEE}%
        \vspace{-0.9cm}
        \thanks{This research was supported by the Swiss National Science
                Foundation under NCCR Automation, grant agreement 
                51NF40\_180545.}%
        \thanks{M.\,A.\,Desai, F.\,Escobar and G.\,Hug are with the Power
        Systems Laboratory at ETH Zurich, 8092 Zurich, Switzerland. Emails:
        {\{desai, escobarprado, hug\}@eeh.ee.ethz.ch}. \textit{(Corresponding
        Author: M.\,A.\,Desai)}}%
}%
\fi

\begin{document}

\begingroup
\allowdisplaybreaks

\maketitle

\begin{abstract}
    Coordinate transformations significantly simplify power systems
    computations. Most notably, the classical Clarke and $dq0$~transformations
    are widely used in three-phase systems, as together they transform balanced
    $abc$~quantities into constant-valued signals. However, during unbalanced
    operation, the utility of these transformations diminishes, since a null
    $0$-coordinate cannot be ensured and oscillating signals emerge. While
    recently proposed transformations ensure a null $0$-coordinate, they \update{either do not lead to constant-valued signals in the $dq0$~domain or fail under various unbalanced scenarios}. In this \update{paper},
    we propose a \update{\gls{GVL}} transformation that ensures both a
    null $0$-coordinate and constant-valued signals. Moreover, we show that, in
    the balanced case, the classical amplitude-invariant Clarke transformation
    is an instance of the proposed \update{\gls{GVL}} transformation.
\end{abstract}
\glsresetall


\vspace{-3mm}
\section{Introduction}
\label{sec:introduction}

Coordinate transformations are crucial for the modeling, analysis, and control
of power systems. Two notable examples~are the Clarke transformation, which
maps $abc$ coordinates to a stationary $\alpha\beta0$~reference frame, and the
$dq0$~transformation, which maps $\alpha\beta0$~coordinates to a rotating
$dq0$~reference frame~\cite{GeometricTransform_Rourke_2019}. In the Clarke
transformation, the basis vectors are chosen \secondupdate{such} that a
balanced set of three-phase quantities produces $\alpha$-
and~$\beta$-coordinates of equal amplitude and in quadrature with each other.
In the $dq0$~transformation, the speed of rotation is set equal to the
electrical frequency,~in which case the $\alpha$- and~$\beta$-coordinates
further simplify to constant-valued signals. Both transformations produce a
null $0$-coordinate in the balanced case. Among other advantages, these
simplifications enable the control of balanced three-phase systems using
elementary techniques, \eg proportional-integral~control.

In an \emph{unbalanced} four-wire system, however, the same choice of basis
vectors for the Clarke transformation fails to achieve the desired
simplifications. This \secondupdate{occurs} because the locus traced out by the
three-phase quantities in Cartesian space morphs from a circle in a canonical
plane to an ellipse in a different plane, \update{which is} determined by the
degree of unbalance.

Recently, transformations have been proposed to address the aforementioned
drawbacks of classical coordinate transformations. In this \update{paper}, we
restrict our focus to transformations that rely on the locus of the space
vector. In~\cite{MNO_Montanari_2017}, a stationary $mno$ reference frame is
proposed, whose orientation depends on the voltage unbalance. However, this
frame is not defined for voltage conditions such as zero voltage
\update{amplitude}
in phase~$a$.
In~\cite{TanSun_3Ph4W_2021}, a non-orthogonal reference frame was proposed for
three-phase four-wire systems; however, when any of the phase
\update{amplitudes}
is
null or the phase difference between any two of them is $0$ or $\pi$ radians,
this transformation fails. The \gls{RRF} proposed
in~\cite{RRF_FMachado_2020} takes advantage of the locus geometry by rotating
and orienting the reference frame \rev{as unit vectors} along the semi-major and semi-minor axes. As
a result, dimensional reduction is obtained in the signals; nonetheless,
constant-valued signals are not obtained by further application of the $dq0$
transformation. Finally, a generalization of the Park transformation has been
proposed in~\cite{Frenet_Milano_2023} using the Frenet-Serret formulae, but, it
also fails to obtain constant-valued quantities.

In this \update{paper}, we propose a \update{\gls{GVL}} transformation that
\rev{overcomes the shortcomings of~the~aforementioned transformations and} is
valid for any type of~unbalance in a three-phase, four-wire system. The
proposed transformation maps the three-phase quantities to two sinusoids of
unit amplitude, in quadrature with each other. Thus,~subsequent application of
the $dq0$~transformation results in two constant-valued signals. We also show
that, in the balanced case, the proposed transformation simplifies to the
classical amplitude-invariant Clarke transformation up to a constant scalar
factor.

\section{\update{Generalized} Vector Locus Transformation}

We consider the three-dimensional space~$\reals[3]$ with orthogonal axes~$abc$
and canonical basis vectors $\unitbasisvector a = \threedvector 100$\!,
$\unitbasisvector b = \threedvector 010$\!, and $\unitbasisvector c =
    \threedvector 001$\!. Inside this space, any set of three-phase quantities can
be represented as a \secondupdate{space} vector of the form $ \spacevector{} =
    \component a\,\unitbasisvector a + \component b\,\unitbasisvector b +
    \component c\,\unitbasisvector c $. For illustration purposes, we
consider~$\spacevector{}$ to be a
\update{vector of phase-to-neutral voltages},
but it can represent other three-phase quantities, such as currents or magnetic
fluxes. In the general case, the individual phase coordinates are
\begin{equation}
    \label{eq:phase-quantities}
    \begin{split}
        \component a
         & = \wave{\amplitude[\verythinspaceneg]{a}}{+\phaseshift a}, \\
        \component b
         & = \wave{\amplitude{b}}{+\phaseshift b - 2\pi/3},           \\
        \component c
         & = \wave{\amplitude{c}}{+\phaseshift c + 2\pi/3},           \\
    \end{split}
\end{equation}
where $\omega$ denotes the fundamental angular frequency.
\secondupdate{Rearranging the coordinates, the voltage vector can be written as}
\begin{equation}
    \label{eq:ellipse}
    \spacevector{}
    =
    \underbrace{\wave[c]{}{-\phaseshift\orientation}}_{\rev{\component 1}}
    \basisvector 1
    +
    \underbrace{\wave[s]{}{-\phaseshift\orientation}}_{\rev{\component 2}}
    \basisvector 2\,,
\end{equation}
where $\phaseshift\orientation\update{\in\left[0,2\pi\right)}$ is an arbitrary angle and the constant vectors
$\basisvector 1$ and~$\basisvector 2$ are given by
\begin{equation}
    \begin{split}
        \label{eq:e1-e2}
        \basisvector 1
        =
        \evaluateat{\spacevector{}}{\sinusoidargument=\phaseshift\orientation}
        =
        \threedvector[l]
        {\sinusoid{c}{\amplitude[\verythinspaceneg]a}{
                \phaseshift\orientation
                +
                \phaseshift[\verythinspaceneg] a
            }}
        {\sinusoid{c}{\amplitude b}{
                \phaseshift\orientation
                +
                \phaseshift b
                -
                2\pi/3
            }}
        {\sinusoid{c}{\amplitude c}{
                \phaseshift\orientation
                +
                \phaseshift c
                +
                2\pi/3
            }} & ,   \\
        \basisvector 2
        =
        \evaluateat{\spacevector{}}{\sinusoidargument=\phaseshift\orientation+\pi/2}
        =
        - \threedvector[l]
          {\sinusoid{s}{\amplitude[\verythinspaceneg]a}{
                  \phaseshift\orientation
                  +
                  \phaseshift[\verythinspaceneg] a
              }}
          {\sinusoid{s}{\amplitude b}{
                  \phaseshift\orientation
                  +
                  \phaseshift b
                  -
                  2\pi/3
              }}
          {\sinusoid{s}{\amplitude c}{
                  \phaseshift\orientation
                  +
                  \phaseshift c
                  +
                  2\pi/3
              }} & . \\
    \end{split}
\end{equation}

In the balanced case, where $\amplitude a = \amplitude b = \amplitude c = V$
and $\phaseshift a = \phaseshift b = \phaseshift c$, we have
$\mynorm{\basisvector 1} = \mynorm{\basisvector 2} = \sqrt{3/2}\,V$\!. Thus,
\eqref{eq:ellipse} reduces to the parametric equation of a circle centered at
the origin with a radius of $\sqrt{3/2}\,V$\!. The amplitude-invariant Clarke
transformation maps the canonical basis vectors $\unitbasisvector a$,
$\unitbasisvector b$, and $\unitbasisvector c$ to $\basisvector{\alpha} =
    \threedvector{1}{-1/2}{-1/2}$, $\basisvector {\beta} =
    \threedvector{0}{\sqrt{3}/2}{-\sqrt{3}/2}$, and $\basisvector{0} =
    \threedvector{1}{1}{1}$, respectively. This results in two perpendicular basis
vectors coplanar with the balanced voltage vector locus and one basis vector
perpendicular to this plane. Therefore, during balanced operation, the Clarke
transformation reduces the three original signals to two displaced by a quarter
of the period, and further application of the $dq0$ transformation
\update{yields}
constant-valued signals.
However, in the general case,~\eqref{eq:ellipse} represents the parametric
equation of an ellipse centered at~the origin. Thus,
\update{under unbalance},
the Clarke transformation does not achieve the objectives mentioned
above, as shown in \cref{fig:classical-clarke}.

\def\VA{1.00}%
\def\VB{1.00}%
\def\VC{1.00}%
\def\phiAdegs{+0.00}%
\def\phiBdegs{+0.00}%
\def\phiCdegs{+0.00}%
\def\phiArads{+0.00}%
\def\phiBrads{+0.00}%
\def\phiCrads{+0.00}%
\def\phiOFFSETdegs{-50.00}%
\def\fHz{+50.00}%
\def\VAU{0.70}%
\def\VBU{1.00}%
\def\VCU{0.40}%
\def\phiAdegsU{-20.00}%
\def\phiBdegsU{+40.00}%
\def\phiCdegsU{-40.00}%
\def\phiAradsU{-0.35}%
\def\phiBradsU{+0.70}%
\def\phiCradsU{-0.70}%
\def\angA{-0.87}%
\def\angB{-2.97}%
\def\angC{+1.22}%
\def\angAU{-1.22}%
\def\angBU{-2.27}%
\def\angCU{+0.52}%
\begin{figure}
    \centering
\begin{tikzpicture}[
	inner frame sep=0,
]


	\newlength\plotwidth\setlength\plotwidth{0.93\columnwidth}
	\newlength\plotheight\setlength\plotheight{0.32\columnwidth}
	
	\newif\ifbalanced\balancedtrue

	\def\wt{360*\fHz*x/1000}
	\def\va{\ifbalanced\VA\else\VAU\fi*cos(\wt + \ifbalanced\phiAdegs\else\phiAdegsU\fi + \phiOFFSETdegs)}
	\def\vb{\ifbalanced\VB\else\VBU\fi*cos(\wt + \ifbalanced\phiBdegs\else\phiBdegsU\fi + \phiOFFSETdegs - 120)}
	\def\vc{\ifbalanced\VC\else\VCU\fi*cos(\wt + \ifbalanced\phiCdegs\else\phiCdegsU\fi + \phiOFFSETdegs + 120)}

	\pgfplotsset{
		x axis line style={opacity=0},
		xtick=\empty,
	}

	\begin{axis}[
		range frame,
		at={(0, 5.75cm)},
		anchor=south,
		width=\plotwidth,
		height=\plotheight,
		xticklabels={},
		xmin=0,
		xmax=40,
		precise y=0,
		ytick={-1, 0, 1},
		clip=false,
	]


		\addplot+[
			curve,
			domain=0:20,
			samples=100,
			colorA
		] {\va}
            node[pos=0.15, above]{$\component[\verythinspaceneg]a$}
		node[hook](A end){};

		\addplot+[
			curve,
			domain=0:20,
			samples=100,
			colorB
		] {\vb}
			node[pos=0.48, above]{$\component b$}
		node[hook](B end){};

		\addplot+[
			curve,
			domain=0:20,
			samples=100,
			colorC
		] {\vc}
					node[pos=0.81, above]{$\component c$}
		node[hook](C end){};

		\balancedfalse

		\addplot+[
			curve,
			domain=20:40,
			samples=100,
			colorA
		] {\va}
        node[right, yshift=-0.15cm]{$\component[\verythinspaceneg] a$}
		node[pos=0, hook](A begin){};

		\addplot+[
			curve,
			domain=20:40,
			samples=100,
			colorB
		] {\vb}
		node[right]{$\component b$}
		node[pos=0, hook](B begin){};

		\addplot+[
			curve,
			domain=20:40,
			samples=100,
			colorC
		] {\vc}
		node[right, yshift=0.05cm]{$\component c$}
		node[pos=0, hook](C begin){};

		\draw[curve, colorA] (A end) -- (A begin);
		\draw[curve, colorB] (B end) -- (B begin);
		\draw[curve, colorC] (C end) -- (C begin);

	\end{axis}


	\path (0, 0) node[inner sep=0pt, anchor=south]{%
		\begin{tikzpicture}[scale=0.95]%
		\input{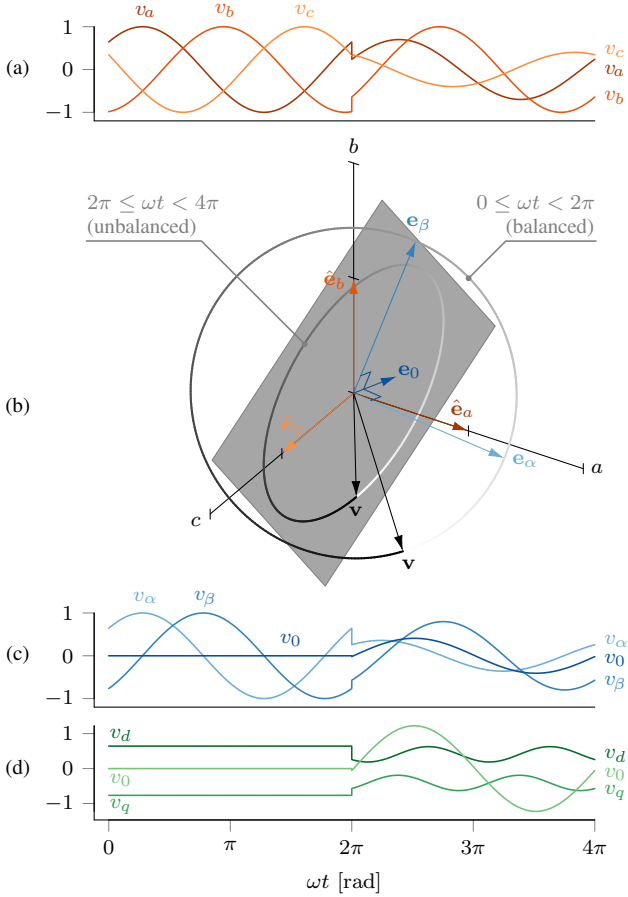}

				\draw[callout] 
					(45:2.16) coordinate (temp) 
					-- 
					++(45:0.7) 
					-- 
					++(0:1.2) 
					node[above left, yshift=0.05cm, align=right]
					{$0\leq \sinusoidargument<2\pi$\\(balanced)};
				\draw[fill, callout] (temp) circle (0.8pt);
				\draw[callout]
					(135:0.92) coordinate (temp)
					--
					++(135:1.942)
					--
					++(180:1.53)
					node[above right, yshift=0.05cm, align=left]
					{$2\pi\leq\sinusoidargument<4\pi$\\(unbalanced)};
				\draw[fill, callout] (temp) circle (0.8pt);

                \path[draw=none, fill=black!70] (origin) circle(0.5pt);

		\end{tikzpicture}%
	};


	\begin{axis}[
		range frame,
		at={(0, -1.22cm)},
		anchor=south,
		width=\plotwidth,
		height=\plotheight,
		xticklabels={},	
		xmin=0,
		xmax=40,
		precise y=0,
		ytick={-1, 0, 1},
		clip=false,
	]

		\balancedtrue

		\addplot+[
			curve,
			domain=0:20,
			samples=100,
			colorCLARKEALPHA
		] {
			0.66666667 * \va 
			+
		   -0.33333333 * \vb
			+
		   -0.33333333 * \vc
		} 
			node[pos=0.15, above]{$\component\alpha$}
		node[hook] (A end){};

		\addplot+[
			curve,
			domain=0:20,
			samples=100,
			colorCLARKEBETA
		] {
			0.00000000 * \va
			+
		    0.57735027 * \vb
			+
		   -0.57735027 * \vc
		}
			node[pos=0.40, above, yshift=-0.05cm]{$\component[\verythinspaceneg]\beta$}
		node[hook] (B end){};

		\addplot+[
			curve,
			domain=0:20,
			samples=100,
			colorCLARKEZERO
		] {
			0.33333333 * \va
			+
		    0.33333333 * \vb
			+
		    0.33333333 * \vc
		}
			node[pos=0.75, above]{$\component0$}
		node[hook] (C end){};

		\balancedfalse

		\addplot+[
			curve,
			domain=20:40,
			samples=100,
			colorCLARKEALPHA
		] {
			0.66666667 * \va
			+
		   -0.33333333 * \vb
			+
		   -0.33333333 * \vc
		}
		node[yshift=0.05cm, right]{$\component \alpha$}
		node[pos=0, hook](A begin){};

		\addplot+[
			curve,
			domain=20:40,
			samples=100,
			colorCLARKEBETA
		] {
			0.00000000 * \va
			+
		    0.57735027 * \vb
			+
		   -0.57735027 * \vc
		}
		node[right, yshift=-0.05cm]{$\component[\verythinspaceneg]\beta$}
		node[pos=0, hook](B begin){};

		\addplot+[
			curve,
			domain=20:40,
			samples=100,
			colorCLARKEZERO
		] {
			0.33333333 * \va
			+
		    0.33333333 * \vb
			+
		    0.33333333 * \vc
		}
		node[yshift=-0.05cm, right]{$\component 0$}
		node[pos=0, hook](C begin){};

		\draw[curve, colorCLARKEALPHA] (A end) -- (A begin);
		\draw[curve, colorCLARKEBETA] (B end) -- (B begin);
		\draw[curve, colorCLARKEZERO] (C end) -- (C begin);

	\end{axis}

	\drawxtrue

	\begin{axis}[
		range frame,
		at={(0, -2.75cm)},
		anchor=south,
		width=\plotwidth,
		height=\plotheight,
		xlabel={$\sinusoidargument$},
		x unit={\si{\radian}},
		xticklabels={$0$, $\pi$, $2\pi$, $3\pi$, $4\pi$},	
		xtick={0, 10, 20, 30, 40},
		xmin=0,
		xmax=40,
		precise y=0,
		ytick={-1, 0, 1},
		clip=false,
	]

		\balancedtrue

		\addplot+[
			curve,
			domain=0:20,
			samples=100,
			colorPARKD
		] {
			0.66666667 * cos(\wt) * \va
			+
			0.66666667 * cos(\wt - 120) * \vb
			+
			0.66666667 * cos(\wt + 120) * \vc
		} 
		node[pos=0.05, yshift=0.15cm]{$\component d$}
		node[hook] (A end){};

		\addplot+[
			curve,
			domain=0:20,
			samples=100,
			colorPARKQ
		] {
			-0.66666667 * sin(\wt) * \va
			+
			-0.66666667 * sin(\wt - 120) * \vb
			+
			-0.66666667 * sin(\wt + 120) * \vc
		}
		node[pos=0.05, yshift=-0.15cm]{$\component q$}
		node[hook] (B end){};

		\addplot+[
			curve,
			domain=0:20,
			samples=100,
			colorPARKZERO
		] {
			\va + \vb + \vc
		}
		node[pos=0.05, yshift=-0.15cm]{$\component 0$}
		node[hook] (C end){};

		\balancedfalse

		\addplot+[
			curve,
			domain=20:40,
			samples=100,
			colorPARKD
		] {
			0.66666667 * cos(\wt) * \va
			+
			0.66666667 * cos(\wt - 120) * \vb
			+
			0.66666667 * cos(\wt + 120) * \vc
		}
		node[right, yshift=0.05cm]{$\component d$}
		node[pos=0, hook](A begin){};

		\addplot+[
			curve,
			domain=20:40,
			samples=100,
			colorPARKQ
		] {
			-0.66666667 * sin(\wt) * \va
			+
			-0.66666667 * sin(\wt - 120) * \vb
			+
			-0.66666667 * sin(\wt + 120) * \vc
		}
		node[right, yshift=-0.05cm]{$\component q$}
		node[pos=0, hook](B begin){};

		\addplot+[
			curve,
			domain=20:40,
			samples=100,
			colorPARKZERO
		] {
			\va + \vb + \vc
		}
		node[right, yshift=-0.05cm]{$\component 0$}
		node[pos=0, hook](C begin){};

		\draw[curve, colorPARKD] (A end) -- (A begin);
		\draw[curve, colorPARKQ] (B end) -- (B begin);
		\draw[curve, colorPARKZERO] (C end) -- (C begin);

	\end{axis}

	\drawxfalse

	\newlength\mylabelsep
	\setlength\mylabelsep{0.49\columnwidth}

	\path (-\mylabelsep, 6.45cm) node {(a)};
	\path (-\mylabelsep, 2.4cm) node {(b)};
	\path (-\mylabelsep, -0.52cm) node {(c)};
	\path (-\mylabelsep, -2.07cm) node {(d)};

\end{tikzpicture}%
    \vspace{-0.2cm}
    \caption{Clarke and $dq0$ transformations applied to three-phase
        quantities.
        (a)~Coordinates of~$\spacevector{}$ in~$abc$ during balanced and
        unbalanced operation.
        (b)~Corresponding loci of~$\spacevector{}$ in~$\reals[3]$ and
        basis vectors of both $abc$ and $\alpha\beta0$.
        (c)~Coordinates of~$\spacevector{}$ in~$\alpha\beta0$.
        (d)~Coordinates of~$\spacevector{}$ in~$dq0$ for axes that rotate
        at the (synchronous) electrical frequency.}
    \label{fig:classical-clarke}%
    \vspace{-0.3cm}
\end{figure}

The reformulated voltage vector as expressed in \eqref{eq:ellipse} facilitates
the realization of the intended \update{\gls{GVL}} transformation. \rev{We
    observe that, for any choice of $\phaseshift\orientation$, the vectors
    $\basisvector 1$ and~$\basisvector 2$ are the \secondupdate{conjugate
        semi-diameters} of the elliptical locus of the voltage vector during unbalanced
    operation.} Thus, we define a third vector $\basisvector 3$ that is
perpendicular to this plane~as
\begin{equation}
    \label{eq:e0}
    \basisvector 3
    =
    (\basisvector 1 \times \basisvector 2)
    \, / 
    \lambda
    \,,
\end{equation}
where the scaling factor $\lambda=\mynorm{\basisvector 1 \times\basisvector 2}/\sqrt{3}$ \update{allows backward compatibility with the classical Clarke transformation in the balanced case}. \rev{\update{Note that} the vector $\basisvector 3$ ensures that the $3$-coordinate $\component 3$ is always null.} Consequently, we propose a transformation such that the canonical
basis vectors $\unitbasisvector{a}$, $\unitbasisvector{b}$,
and $\unitbasisvector{c}$ are transformed into $\basisvector1$,
$\basisvector2$, and $\basisvector3$. The inverse of the proposed transformation can therefore be represented as
\begin{equation}
    \threedvector{\component a}{\component b}{\component c}
    =
    \underbracedmatrix{
        \matrixwithcols{\basisvector 1}{\basisvector 2}{\basisvector 3}
    }{
        \GVLMatrix[i]
    }
    \threedvector{\component 1}{\component 2}{\component 3}.
\end{equation}
The proposed \update{\gls{GVL}} transformation $\GVLMatrix$ reduces the three
original unbalanced signals to two sinusoidal signals displaced by a quarter of
the period. \update{Moreover,
    the two signals have the same (unit) amplitude. \secondupdate{This is enabled
        by the form of~$\basisvector 1$~and~$\basisvector2$ in~\eqref{eq:e1-e2}, which
        retains the essential information about the unbalance (in contrast to
        normalized vectors).} The further application of the $dq0$ transformation can
    also be visualized as rotating the
    vectors $\basisvector 1$ and $\basisvector 2$ such that, for every time
    instant, the corresponding \secondupdate{conjugate semi-diameters} of the
    ellipse are obtained. Thus, this application leads to two
    constant-valued~signals.}

\update{At first glance, we note that~$\GVLMatrix$ is written in terms of the unbalanced parameters.
    \secondupdate{According to~\cite{TanSun_3Ph4W_2022}, these can be estimated.}
    However, the key observation is that $\basisvector1$ and $\basisvector2$ are samples of $\spacevector{}$ separated by a quarter of the period and
    $\basisvector3$ is obtained using~\eqref{eq:e0}. This observation can be leveraged in closed-loop implementations by using a frequency estimator to enforce the quarter period separation without using an explicit delayed measurement.}


The instant used to compute the first basis vector is set by the angle
$\phaseshift\orientation$ introduced earlier. \rev{There are infinitely many
    valid choices of $\phaseshift\orientation$, each yielding a different pair of
    conjugate semi-diameters $\basisvector 1$ and $\basisvector 2$; we discuss
    two.} The first choice is the value
\update{$\classical{\phaseshift\orientation}=-\phaseshift{a}$},
which corresponds
to measuring~$\spacevector{}$ when $\component a$ reaches its maximum. In the
balanced case, this leads to the simplification of~$\GVLMatrix$ to the
amplitude-invariant Clarke transformation up to a constant factor that ensures
coordinates with unit amplitude. The second choice is the value
\update{$\desired{\phaseshift\orientation}=\arg\max\mynorm{\spacevector{}} = \arg\max{\mynorm{\spacevector{}}}^2$}.
Since
${\mynorm{\spacevector{}}}^2 = \component a^2 + \component b^2 + \component c^2$,
using trigonometric identities, it can be succinctly represented in the form
${\mynorm{\spacevector{}}}^2 = C-\doubleanglewave{A}{+\auxangle}$
for
\secondupdate{constants}
$C$, $A$, and~$\auxangle$.
Thus, $\desired{\phaseshift\orientation}= -\pi/4 - \auxangle/2$, where
\begin{equation}
    \tan\auxangle
    =
    -
    \frac{
        \begin{multlined}
            \sinusoid[n] c{\amplitude a^2}{2\phaseshift a}
            +
            \sinusoid c{\amplitude b^2}{2\phaseshift b - 4\pi/3}
                + {} \\
            \sinusoid c{\amplitude c^2}{2\phaseshift c + 4\pi/3}
        \end{multlined}
    }{
        \begin{multlined}
            \sinusoid[n] s{\amplitude a^2}{2\phaseshift a}
            +
            \sinusoid s{\amplitude b^2}{2\phaseshift b - 4\pi/3}
                + {} \\
            \sinusoid s{\amplitude c^2}{2\phaseshift c + 4\pi/3}
        \end{multlined}
    }\,.
\end{equation}
In this case, we obtain an orthogonal set of basis vectors such that
$\basisvector 1$ and~$\basisvector 2$ are aligned with the major and minor axes
of the ellipse, respectively, which simplifies the instantaneous power calculation \update{(see \cref{sec:power-calculations})}.
Normalizing $\basisvector 1$, $\basisvector2$, and~$\basisvector 3$ for
this~$\desired{\phaseshift\orientation}$ recovers the \gls{RRF}
of~\cite{RRF_FMachado_2020}. \cref{fig:new-basis-vectors}
\secondupdate{gives}
the geometric interpretation of the special
choices.

\begin{figure}
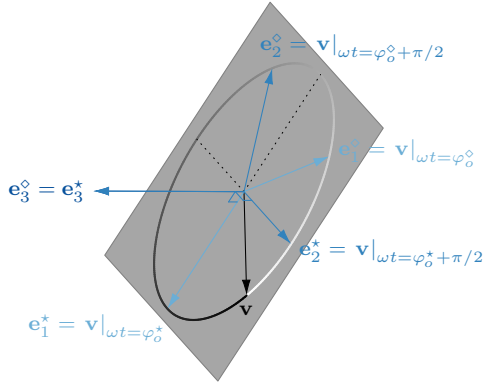

    \centering
    \input{figures/new_basis_vectors_parameters_auto}%
    \begin{tikzpicture}


	\path (0, 0) node[inner sep=0pt, anchor=south] {%
		\begin{tikzpicture}[scale=0.82]%
			\input{figures/new_basis_vectors_3D}
		\end{tikzpicture}%
	};


\end{tikzpicture}%
    \vspace{-0.1cm}
    \caption{Special choices of the basis vectors, found by measuring
        $\spacevector{}$ when $\component[\verythinspaceneg] a$ reaches
        its peak ($\diamond$) or $\mynorm{\spacevector{}}$ is maximized
        ($\star$). The vector~$\basisvector 3$, being orthogonal to the
        ellipse and having length~$\sqrt 3$, is independent of these
        choices.}
    \label{fig:new-basis-vectors}
    \vspace{-0.5cm}
\end{figure}

\rev{We note that \eqref{eq:ellipse}
    \secondupdate{describes}
    a linear locus, and the matrix $\GVLMatrix[i]$
    \secondupdate{becomes}
    non-invertible, \emph{only} \update{if} \update{all three} $abc$ coordinates
    \secondupdate{are}
    linearly dependent.}
\update{Specifically, this linear dependence can appear if (i) all $abc$ coordinates are in phase with each other, (ii) at least two of their amplitudes are~$0$, or (iii) the amplitude of one coordinate is~$0$ \emph{and} the phase difference between the other two coordinates is $0$ or $\pi$ radians.}
\update{These particular unbalanced scenarios,}
where only one
effective signal exists instead of three,
\secondupdate{renders}
further dimensionality
reduction from one to two signals illogical. Such an attempt
\secondupdate{does}
not decrease
the dimensionality, but rather introduce potential inaccuracies and redundancy
by trying to add an unsupported dimension.

\update{A comparison with the other transformations \secondupdate{discussed in \cref{sec:introduction} is presented} in \cref{tab:comparison}. In summary, the proposed \gls{GVL} is the only transformation that ensures equal-amplitude, in-quadrature outputs and constant $dq0$ components under unbalance. \camera{Moreover, its sole failure mode is the genuinely degenerate case in which the voltage locus collapses to a line, a limitation shared by all such transformations; the $mno$ and Tan-Sun frames, in contrast, also fail under conditions that leave the locus non-degenerate, namely a single $0$ phase amplitude or any two phase angles differing by $0$ or~$\pi$.}}

\begin{table*}
    \newcommand\abc{abc}
    \newcommand\abz{\alpha\beta0}
    \newcommand\dqz{dq0}
    \newcommand\maps[2]{$#1\mapsto#2$}
    \newcommand\oscil{Oscillatory}
    \newcommand\linearlocus{Locus of\,$\spacevector{}$\,is linear}
    \newcommand\zeroamp{Any amplitude is~$0$}
    \centering
    \begin{threeparttable}
        \caption{Comparison of the proposed
            \update{\gls{GVL}}
            transformation against existing
            transformations
        }
        \label{tab:comparison}
        \begin{tabularx}\textwidth{llllll}
            \toprule
                           &                                       & \multicolumn{2}{l}{Output coords.\,under unbalance} &               &                                                                                                      \\\cmidrule(lr){3-4}
            Transformation & Generalizes                           & Equal amplitude                                     & In quadrature & Resulting $\dqz$ coords.\,under unbalance & Failure scenarios \secondupdate{(non-invertible matrix)} \\
            \midrule
            $mno$~\cite{MNO_Montanari_2017}
                           & \maps\abc\abz
                           & No                                    & No
                           & \oscil
                           & \linearlocus; \MakeLowercase\zeroamp.
            \\
            Tan-Sun~\cite{TanSun_3Ph4W_2021,TanSun_3Ph4W_2022}
                           & \maps\abc\abz
                           & Yes                                   & Yes
                           & Constant
                           & \linearlocus; \MakeLowercase\zeroamp;
            \\
                           &                                       &                                                     &               &                                           & any two phase angles differ by~$0$ or $\pi$.
            \\[0.75mm]
            \gls{RRF}~\cite{RRF_FMachado_2020}
                           & \maps\abc\abz
                           & No                                    & Yes
                           & \oscil
                           & \linearlocus.\,\tnote{a}
            \\
            Frenet frame~\cite{Frenet_Milano_2023}
                           & \maps\abc\dqz
                           & ---                                   & ---
                           & \oscil\,\tnote{b}
                           & \linearlocus.
            \\
            \gls{GVL} (proposed)
                           & \maps\abc\abz
                           & Yes\,\tnote{c}\!\!\!                  & Yes
                           & Constant
                           & \linearlocus.
            \\
            \bottomrule
        \end{tabularx}
        \begin{tablenotes}
            \update{\item[a] An additional check is implemented for the linear locus case and a new
                matrix is constructed as the original \gls{RRF} transformation matrix is
                non-invertible.}
            \item[b] While the so-called~\cite{Frenet_Milano_2023} N- and B-coordinates
            (equivalent to $q$ and $0$, respectively) are both null, the T-coordinate
            (equivalent to $d$) is oscillatory.
            \item[c] The common amplitude is~$1$.
        \end{tablenotes}
    \end{threeparttable}
    \vspace{-0.55cm}
\end{table*}

\section{Power Calculations}
\label{sec:power-calculations}

We consider voltage and current space vectors
${\spacevector{},\,\currentvector{}\in\mathbb{R}^3}$. Without loss of
generality, we note that the proposed transformation maps the canonical basis
$\{\unitbasisvector a,\,\unitbasisvector b,\,\unitbasisvector c\}$ to a basis
$\{\basisvector 1,\,\basisvector 2,\,\basisvector 3\}$, where $\basisvector 1$
and $\basisvector 2$ span the voltage locus plane and $\basisvector 3$ is
normal to it. Under this construction, we have $ \spacevector{} = \component
    1\,\basisvector 1 + \component 2\,\basisvector 2 + 0\,\basisvector 3$ and $
    \currentvector{} = \currentcomponent 1\,\basisvector 1 + \currentcomponent
    2\,\basisvector 2 + \currentcomponent 3\,\basisvector 3$.

The instantaneous active power is given by the dot product of the voltage and
the current space vectors~\update{\cite[\S3.3.6]{akagi2017}}, \ie
\begin{equation}
    \label{eq:activepower}
    \begin{split}
        p & = \spacevector{}\cdot\currentvector{} \\
          & =
        {\voltagecomponent 1}
        {\currentcomponent 1}
        \mynorm{\basisvector 1}^{2}
        +
        {\voltagecomponent 2}
        {\currentcomponent 2}
        \mynorm{\basisvector 2}^{2}
        +
        (
        {\voltagecomponent 1}{\currentcomponent 2}
        +
        {\voltagecomponent 2}{\currentcomponent 1}
        )
        \,
        \basisvector 1\cdot\basisvector 2
        \,.
    \end{split}
\end{equation}
If
\secondupdate{%
    $\phaseshift\orientation=\desired{\phaseshift\orientation}$ is used to form the basis vectors%
},
the~active~power
simplifies to $ p= {\voltagecomponent 1}{\currentcomponent 1}
    \mynorm{\basisvector 1}^{2}\! +{\voltagecomponent 2}{\currentcomponent
        2}\mynorm{\basisvector 2}^{2}\!$, \secondupdate{since
    $\basisvector1\cdot\basisvector2=0$}. This generalizes the usual $\alpha\beta$
dot product, with
scaling factors~equal to the squared lengths of the semi-major and semi-minor
axes.

According to \cite{fangzhengpeng1996}, the instantaneous reactive power is
defined as the cross product of the voltage and the current space vectors,
i.e., $\qvector{} = \voltagevector{}\times\currentvector{}$.
Thus,
\begin{equation}
    \begin{split}
        \label{eq:reactivepower}
        \qvector{}
        =
        (1/\lambda)\,
        (
        \voltagecomponent 2\currentcomponent3
        \mynorm{\basisvector 2}^2
        +
        \voltagecomponent 1\currentcomponent 3
        \,
        \basisvector 1\cdot \basisvector 2
        )
        \,
        \basisvector 1 &
        \\
        -
        \,
        (1/\lambda)\,
        (
        \voltagecomponent 1\currentcomponent 3
        \mynorm{\basisvector 1}^2
        +
        \voltagecomponent 2\currentcomponent 3
        \,
        \basisvector 1 \cdot \basisvector 2
        )
        \,
        \basisvector 2 &
        \\
        +
        \,
        \lambda\,
        (
        \voltagecomponent1\currentcomponent2
        -
        \voltagecomponent2\currentcomponent1
        )
        \,
        \basisvector 3 & \,.
    \end{split}
\end{equation}
Using $\desired{\phaseshift\orientation}$
\secondupdate{in}
the basis vectors,
this simplifies to
\begin{equation}
    \begin{split}
        \label{eq:reactivepower_simp}
        \qvector{}=\sqrt{3}\,{\voltagecomponent 2}{\currentcomponent 3}
        \mynorm{\basisvector 2}
        {\unitbasisvector 1}-\sqrt{3}\,{\voltagecomponent 1}{\currentcomponent 3}
        \mynorm{\basisvector 1}
        {\unitbasisvector 2} &        \\
        +\,({\voltagecomponent 1}{\currentcomponent 2}-{\voltagecomponent 2}{\currentcomponent 1})
        \mynorm{\basisvector 1}\mynorm{\basisvector 2}
        \unitbasisvector 3      & \,,
    \end{split}
\end{equation}
where $\unitbasisvector 1,\,\unitbasisvector 2,\,\unitbasisvector 3$ are the respective unit vectors. It should be noted here that the third term in \eqref{eq:reactivepower}
\secondupdate{and}
\eqref{eq:reactivepower_simp} is the component of the reactive power perpendicular to the voltage plane, analogous to the $\alpha\beta$ reactive power term. In contrast, the first two terms are components in the voltage plane and arise only when the out-of-plane current component is nonzero, which occurs due to the presence of the zero-sequence current.

\section{Numerical Validation}

To validate the proposed \update{\gls{GVL}} transformation, we \update{first}
apply it to the unbalanced system of
\cref{fig:classical-clarke,fig:new-basis-vectors} (interval
${2\pi\leq\sinusoidargument<4\pi}$). The specific $abc$~coordinates are given
by
\begin{equation*}
    \label{eq:phase-quantities-numeric}
    \begin{split}
        \component a
         & = \wave{\VAU}{-7\pi/18},       \\
        \component b
         & = \wave{\VBU}{-\pi/18-2\pi/3}, \\
        \component c
         & = \wave{\VCU}{-\pi/2+2\pi/3},  \\
    \end{split}
\end{equation*}
and are plotted again in \cref{fig:numerical-example}(a).
We consider both choices
$\phaseshift\orientation=\classical{\phaseshift\orientation}$ and
$\phaseshift\orientation=\desired{\phaseshift\orientation}$, with the resulting
coordinates shown in~\cref{fig:numerical-example}(c). The phase shift between,
\eg $\classical{\component1}$ and~$\desired{\component1}$, reflects the time
for~$\spacevector{}$ to traverse the arc between $\classical{\basisvector1}$
and~$\desired{\basisvector1}$ in~\cref{fig:numerical-example}(b). As expected,
the $1$- and $2$-coordinates have unit amplitude and are in quadrature, while
the $3$-coordinate is null throughout. Applying the standard~$dq0$
transformation then yields the constant coordinates
of~\cref{fig:numerical-example}(d), which could readily follow setpoints, \eg
via proportional-integral controllers. \camera{Moreover, since these signals
    are constant, zero-mean measurement noise can be attenuated by low-pass
    filtering without distorting them, unlike for transformations whose $dq0$
    components remain oscillatory.}

\begin{figure}
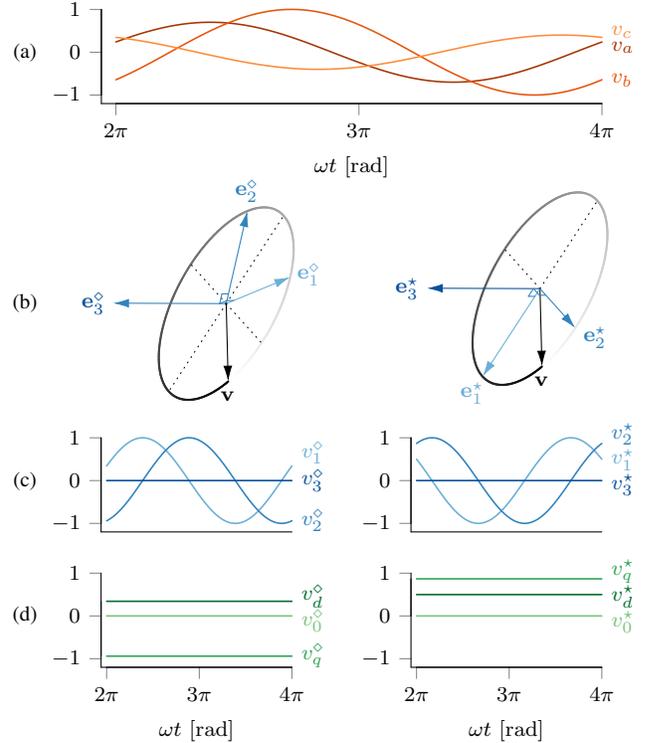

    \centering
    \evaluatefalse%
\begin{tikzpicture}[
	inner frame sep=0
]
    \def\smallsep{2.06cm}

	\setlength\plotwidth{0.93\columnwidth}
	\setlength\plotheight{0.32\columnwidth}

	\newif\ifbalanced\balancedtrue

	\def\wt{360*\fHz*x/1000}
	\def\va{\ifbalanced\VA\else\VAU\fi*cos(\wt + \ifbalanced\phiAdegs\else\phiAdegsU\fi + \phiOFFSETdegs)}
	\def\vb{\ifbalanced\VB\else\VBU\fi*cos(\wt + \ifbalanced\phiBdegs\else\phiBdegsU\fi + \phiOFFSETdegs - 120)}
	\def\vc{\ifbalanced\VC\else\VCU\fi*cos(\wt + \ifbalanced\phiCdegs\else\phiCdegsU\fi + \phiOFFSETdegs + 120)}

	\pgfplotsset{
		x axis line style={opacity=0},
		xtick=\empty,
	}

    \drawxtrue
	\begin{axis}[
		range frame,
		at={(0, 3.7cm)},
		anchor=south,
		width=\plotwidth,
		height=\plotheight,
		xlabel={$\sinusoidargument$},
		x unit={\si{\radian}},
		xticklabels={$2\pi$, $3\pi$, $4\pi$},	
		xtick={0, 10, 20},
		xmin=0,
		xmax=20,
		precise y=0,
		ytick={-1, 0, 1},
		clip=false,
	]


		\balancedfalse

		\addplot+[
			curve,
			domain=0:20,
			samples=200,
			colorA
		] {\va}
		node[right, yshift=-0.07cm]{$\component a$};

		\addplot+[
			curve,
			domain=0:20,
			samples=200,
			colorB
		] {\vb}
		node[right]{$\component b$};

		\addplot+[
			curve,
			domain=0:20,
			samples=200,
			colorC
		] {\vc}
		node[right, yshift=0.07cm]{$\component c$};

	\end{axis}
    \drawxfalse


    \def\locussep{2}
	\path (-\locussep, 0) node[inner sep=0pt, anchor=south]{%
		\begin{tikzpicture}[scale=0.96]%
            \input{figures/new_basis_vectors_3D_classical.tex}
		\end{tikzpicture}%
	};

	\path (\locussep, 0) node[inner sep=0pt, anchor=south]{%
		\begin{tikzpicture}[scale=0.96]%
            \input{figures/new_basis_vectors_3D_desired.tex}
		\end{tikzpicture}%
	};


	\begin{axis}[
		range frame,
        at={(-\smallsep, -1.45cm)},
		anchor=south,
		width=\plotwidth/2,
		height=\plotheight,
		xticklabels={},	
		precise y=0,
		ytick={-1, 0, 1},
		clip=false,
	]

		\balancedfalse

		\addplot+[
			curve,
			colorCLARKEALPHA
        ] table[x=t, y=V1, col sep=comma] {figures/V_123_classical.csv}
        node[yshift=0.17cm, right]{$\classical{\component1}$};

		\addplot+[
			curve,
			colorCLARKEBETA
        ] table[x=t, y=V2, col sep=comma] {figures/V_123_classical.csv}
        node[yshift=-0.0cm, right]{$\classical{\component2}$};

		\addplot+[
			curve,
			colorCLARKEZERO
        ] table[x=t, y=V3, col sep=comma] {figures/V_123_classical.csv}
        node[yshift=-0.0cm, right]{$\classical{\component3}$};

	\end{axis}

	\begin{axis}[
		range frame,
        at={(\smallsep, -1.45cm)},
		anchor=south,
		width=\plotwidth/2,
		height=\plotheight,
		xticklabels={},	
		precise y=0,
		ytick={-1, 0, 1},
		clip=false,
	]

		\balancedfalse

		\addplot+[
			curve,
			colorCLARKEALPHA
        ] table[x=t, y=V1, col sep=comma] {figures/V_123_desired.csv}
        node[yshift=-0.005cm, right]{$\desired{\component1}$};

		\addplot+[
			curve,
			colorCLARKEBETA
        ] table[x=t, y=V2, col sep=comma] {figures/V_123_desired.csv}
        node[yshift=0.125cm, right]{$\desired{\component2}$};

		\addplot+[
			curve,
			colorCLARKEZERO
        ] table[x=t, y=V3, col sep=comma] {figures/V_123_desired.csv}
        node[yshift=-0.06cm, right]{$\desired{\component3}$};

	\end{axis}

	\drawxtrue

	\begin{axis}[
		range frame,
        at={(-\smallsep, -3.05cm)},
		anchor=south,
		width=\plotwidth/2,
		height=\plotheight,
		xlabel={$\sinusoidargument$},
		x unit={\si{\radian}},
		xticklabels={$2\pi$, $3\pi$, $4\pi$},	
		xtick={0, 10, 20},
		precise y=0,
		ytick={-1, 0, 1},
        ymin=-1,
        ymax=1,
		clip=false,
	]

		\balancedfalse

		\addplot+[
			curve,
			colorPARKD
        ] table[x=t, y=Vd, col sep=comma] {figures/V_dq0_classical.csv}
        node[yshift=0.07cm, right]{$\classical{\component d}$};

		\addplot+[
			curve,
			colorPARKQ
        ] table[x=t, y=Vq, col sep=comma] {figures/V_dq0_classical.csv}
        node[yshift=-0.0cm, right]{$\classical{\component q}$};

		\addplot+[
			curve,
			colorPARKZERO
        ] table[x=t, y=V0, col sep=comma] {figures/V_dq0_classical.csv}
        node[yshift=-0.07cm, right]{$\classical{\component 0}$};

	\end{axis}

	\begin{axis}[
		range frame,
        at={(\smallsep, -3.05cm)},
		anchor=south,
		width=\plotwidth/2,
		height=\plotheight,
		xlabel={$\sinusoidargument$},
		x unit={\si{\radian}},
		xticklabels={$2\pi$, $3\pi$, $4\pi$},	
		xtick={0, 10, 20},
		precise y=0,
		ytick={-1, 0, 1},
        ymin=-1,
        ymax=1,
		clip=false,
	]

		\balancedfalse

		\addplot+[
			curve,
			colorPARKD
        ] table[x=t, y=Vd, col sep=comma] {figures/V_dq0_desired.csv}
        node[yshift=-0.03cm, right]{$\desired{\component d}$};

		\addplot+[
			curve,
			colorPARKQ
        ] table[x=t, y=Vq, col sep=comma] {figures/V_dq0_desired.csv}
        node[yshift=0.1cm, right]{$\desired{\component q}$};

		\addplot+[
			curve,
			colorPARKZERO
        ] table[x=t, y=V0, col sep=comma] {figures/V_dq0_desired.csv}
        node[yshift=-0.08cm, right]{$\desired{\component 0}$};

	\end{axis}

	\drawxfalse

	\setlength\mylabelsep{0.49\columnwidth}

	\path (-\mylabelsep, 4.38cm) node {(a)};
	\path (-\mylabelsep, 1.35cm) node {(b)};
	\path (-\mylabelsep, -0.77cm) node {(c)};
	\path (-\mylabelsep, -2.36cm) node {(d)};

\end{tikzpicture}
    \vspace{-0.19cm}
    \caption{Proposed transformation applied to an unbalanced three-phase
        system.
        (a)~Coordinates of~$\spacevector{}$ in $abc$.
        (b)~Choices of the basis vectors.
        (c)~Corresponding coordinates of~$\spacevector{}$ under the new
        transformation.
        (d)~Coordinates of~$\spacevector{}$ in~$dq0$ for axes that rotate
        at the (synchronous) electrical frequency.}
    \vspace{-0.6cm}
    \label{fig:numerical-example}
\end{figure}

\newcommand\makeinterval[2]{#1\leq\omega t < #2}
We next test the proposed
transformation under the more severe unbalances of
\cref{fig:second-numerical-example}(a). \update{Here, we test the cases where
    (\textsc{i}) any amplitude is 0, (\textsc{ii}) the phase difference between any
    two phases is 0, and (\textsc{iii}) the phase difference between any two phases
    is $\pi$.} \camera{The proposed transformation handles all three without any
    special-case detection.} Specifically, we fix $v_a = \cos\omega t$ and $v_c =
    \cos\left(\omega t+2\pi/3\right)$ for $\makeinterval{0}{6\pi}$, and then employ
\begin{alignat*}{3}
     & v_b = 0
     &                                          & \text{ and }
    \phaseshift\orientation =
    0
     &                                          & \text{ for }
    \makeinterval{0}{2\pi} \,,                                 \\
     & v_b = 0.3\cos\omega t\,
     &                                          & \text{ and }
    \phaseshift\orientation =
    2\pi
     &                                          & \text{ for }
    \makeinterval{2\pi}{4\pi} \,,                              \\
     & v_b = 0.3\cos\left(\omega t + \pi\right)
     &                                          & \text{ and }
    \phaseshift\orientation =
    4\pi
     &                                          & \text{ for }
    \makeinterval{4\pi}{6\pi} \,.
\end{alignat*}
The left plots of \cref{fig:second-numerical-example}(b) and~(c) show the $123$
and $dq0$ coordinates from the proposed \gls{GVL} transformation; the right
plots show the \gls{RRF}~\cite{RRF_FMachado_2020}, which by
\cref{tab:comparison} is the only other $abc$-to-$\alpha\beta0$ generalization
admitting these scenarios. The RRF yields quadrature sinusoids of unequal
amplitude, so further $dq0$ application does not give constant quantities.

\begin{figure}[t!]
    \centering
    \input{figures/parameters.tex}%
    \input{figures/new_example.tex}
    \vspace{-0.2cm}
    \caption{Comparison of the proposed transformation and the \gls{RRF} transformation~\cite{RRF_FMachado_2020} for different, severe degrees of unbalance, namely when $v_b=0$~(\textsc i), when $v_a$ and $v_b$ are in phase~(\textsc{ii}), or when they are in counterphase~(\textsc{iii}).
        (a)~Coordinates of~$\spacevector{}$ in $abc$.
        (b)~Corresponding coordinates of~$\spacevector{}$ under the compared transformations.
        (c)~Coordinates of~$\spacevector{}$ in~$dq0$ for axes that rotate
        at the (synchronous) electrical frequency.}
    \label{fig:second-numerical-example}
    \vspace{-0.5cm}
\end{figure}

\section{Conclusion}

In this \update{paper}, we examined the shortcomings of the classical Clarke
and $dq0$~transformations for the unbalanced operation of a three-phase system.
We also listed the shortcomings of recently proposed transformations for
unbalanced operation. Addressing these, we propose a \update{\gls{GVL}}
transformation, whose effectiveness is substantiated through
\update{numerical examples}.
The proposed
transformation converts any unbalanced three-phase quantities into two
unit-amplitude sinusoids in quadrature, leading to two constant-valued signals
after applying the $dq0$~transformation. In the balanced scenario, it reduces
to the classical Clarke transformation, scaled by a constant. Furthermore, the
straightforward implementation of this \update{\gls{GVL}} transformation, coupled with its
backward compatibility with established classical transformations, advocates
its adoption in the control of three-phase systems. \rev{Future work will explore the closed-loop implementation of the proposed transformation.}

\section*{AI Usage Disclosure}

\update{The authors did \emph{not} use Artificial Intelligence (AI) either in the research or in the writing of this paper.}










\bibliographystyle{IEEEtran}
\bibliography{bibliography.bib}
\balance

\endgroup
\end{document}